\documentclass[11pt,twoside]{article}
\usepackage{asp2006}
\usepackage{epsf}
\usepackage{lscape}

\begin{document}
\title{On the Primordial Helium Abundance and the $\Delta Y/\Delta O $ Ratio}
\author{Manuel Peimbert\altaffilmark{1}, Valentina Luridiana\altaffilmark{2}, 
Antonio Peimbert\altaffilmark{1}, \& Leticia
Carigi\altaffilmark{1}} 

\altaffiltext{1}{Instituto de
Astronom\'{\i}a, Universidad Nacional Aut\'onoma de M\'exico; Apdo. postal
70--264; Ciudad Universitaria; M\'exico D.F. 04510; Mexico.}

\altaffiltext{2}{Instituto de
Astrof\'{\i}sica de Andaluc\'{\i}a (CSIC), Camino bajo de Hu\'etor 50,
18008 Granada, Spain.}

\begin{abstract}

We present a review on the determination of the primordial helium abundance,
$Y_p$, based on the study of hydrogen and helium recombination lines in
extragalactic H~{\sc{ii}} regions. We also discuss the observational 
determinations of the increase of helium to the increase of oxygen
by mass $\Delta Y/\Delta O $, and compare them with predictions based
on models of galactic chemical evolution.
\end{abstract}

\section{Why $Y_p$?}

The determination of $Y_p$ is important for at least the following reasons:
a) It is one of the pillars of Big Bang cosmology and an accurate determination
of $Y_p$ permits to test the standard Big Bang nucleosynthesis (SBBN), b) the models of
stellar evolution require an accurate initial $Y$ value; this is given by $Y_p$
plus the additional $Y$ produced by galactic chemical evolution, which can be estimated
based on the $\Delta Y/\Delta O $ ratio, c) the combination of
$Y_p$ and $\Delta Y/\Delta O $ is needed to test models of galactic chemical
evolution, d) to test solar models it is necessary to know the initial solar abundances,
which are different to the photospheric ones due to diffusive settling, this effect reduces
the helium and heavy element abundances in the solar photosphere  
relative to that of hydrogen, 
the initial solar abundances can be provided by models of galactic chemical 
evolution, e) the determination of the $Y$ value 
in metal poor H~{\sc ii} regions requires a deep knowledge of their physical conditions, 
in particular the $Y$ determination depends to a significant degree on their density and
temperature distribution, therefore accurate $Y$ determinations combined with the assumption of
SBBN provide a constraint on the density and temperature structure of H~{\sc ii} regions.

\section{Recent Determinations of $Y_p$}

Previous reviews on $Y_p$ determinations have been presented by \citet{pei03} and
\citet{lur03}.
Recent determinations of $Y_p$ are those by \citet{izo99}, \citet{izo04},
\citet{izo06}, \citet{lal03}, \citet{oli04}, \citet{fuk06}, and \citet{pei07}. A
critical discussion of these determinations has been presented by
\citet{pei07}. Most of the differences among these determinations are
due to systematic effects.

\section{Error Budget}

The error budget of the $Y_p$ determination by \citet{pei07} is presented in 
Table~1. In this table the sources of error are listed in order
of importance. The error budgets of $Y_p$ determinations by other groups are different
to that by \citet{pei07} for many reasons, each error budget depends on the 
sample of H~{\sc ii} regions used and on the treatment given to the different sources of error.

\begin{table}[!ht]
\label{tab:error}
\caption{Error budget in the $Y_p$ determination}
\begin{center}
{\small
\begin{tabular}{lcc@{$\pm$}r@{}c}
\tableline
\tableline
\noalign{\smallskip}
Problem & \multicolumn{4}{c}{Estimated error}\\
\noalign{\smallskip}
\tableline 
\noalign{\smallskip}
Collisional Excitation of the H~{\sc{i}} Lines         &\hfill&& 0.0015 &\hfill\\
Temperature Structure                                  &&& 0.0010 &\\
$O$ $(\Delta Y/\Delta O)$ Correction                   &&& 0.0010 &\\
Recombination Coefficients of the He~{\sc{i}} Lines    &&& 0.0010 &\\
Density Structure                                      &&& 0.0007 &\\
Underlying Absorption in the He~{\sc{i}} Lines         &&& 0.0007 &\\
Recombination Coefficients of the H~{\sc{i}} Lines     &&& 0.0005 &\\
Underlying Absorption in the H~{\sc{i}} Lines          &&& 0.0005 &\\
Ionization Structure                                   &&& 0.0005 &\\
Collisional Excitation of the He~{\sc{i}} Lines        &&& 0.0005 &\\
Reddening correction                                   &&& 0.0005 &\\
Optical Depth of the He~{\sc{i}} Triplet Lines         &&& 0.0003 &\\
He~{\sc{i}} and H~{\sc{i}} Line Intensities            &&& 0.0003 &\\
\noalign{\smallskip}
\tableline
\end{tabular}
}
\end{center}
\end{table}

\section{The Four Main Sources of Error}

The most important source of error is the collisional 
excitation of Balmer lines. Neutral hydrogen atoms in excited states may form
not only by the usual process of H$^+$ recombination, but also through
collisions of neutral hydrogen with electrons. The recombination cascade that follows
such excitations contributes to the observed intensity of Balmer lines, mimicking
a larger relative hydrogen abundance and, hence, a smaller helium abundance.
To estimate this contribution it is
necessary to have a tailored photoionization model for each object
that fits properly the temperature structure. The contribution to the Balmer
line intensities depends strongly on the temperature: therefore this effect, 
and consequently the associated error in its estimate,
increases for H~{\sc ii} regions of lower metallicity and consequently
higher temperature.

The second most important source of error is the temperature structure. Most
determinations neglect the possible presence of temperature 
variations across the H~{\sc ii} region structure and assume that $T$(O~{\sc iii})
is representative of the zone where the He~{\sc i} recombination lines form.
However, other temperature determinations based on different diagnostics yield lower values;
furthermore, photoionization models do not predict the high
$T$(O~{\sc iii}) values observed. These results indicate that
temperature variations are indeed present in H~{\sc ii} regions, and 
this result should be included in the $Y$ determination \citep{pea02}. 
The best procedure to take into account the temperature
structure is to self-consistently determine 
$T$(He~{\sc ii}) from a set of He~{\sc i} lines by means of the maximum likelihood method. The $Y$ abundances derived 
from $T$(He~{\sc ii}) are typically lower by about 0.0030 $-$ 0.0050 than those 
derived from $T$(O~{\sc iii}). The difference between both temperatures does not have
a significant trend with the metallicity of the H~{\sc ii} region, 
hence the systematic error introduced by the use of $T$(O~{\sc iii}) 
in the $Y$ determination is similar for objects with different metallicities.
The error quoted under ``Temperature Structure'' in Table~1 is
the residual error due to the uncertainty in the $T$(He~{\sc ii})
determinations of the dataset by \citet{pei07}.

The third most important source of error is the extrapolation of the
derived $Y$ values to zero metallicity through the $\Delta Y/\Delta O $
ratio. This problem will be discussed in the next section. The fourth
most important source of error is the uncertainty on the recombination coefficients of the He~{\sc i}
lines.

\section{$\Delta Y/\Delta O $ from Models and Observations}

To determine the $Y_p$ value it is customary to use the $Y$ values of a set of
$O$ poor galaxies and to extrapolate the $Y$ values to the case of 
$O=0$ using the following equation:

\begin{equation}
Y_p  =  Y - { O} \frac{\Delta Y}{\Delta O}
\label{eq:DeltaO}
,\end{equation}
where $O$ is the oxygen abundance per unit mass.
To obtain an accurate $Y_p$ value, a reliable determination of 
$\Delta Y/\Delta O $ for $O$-poor objects is needed.

The $\Delta Y/\Delta O $ value derived by \citet{pei00} from observational
results and models of chemical evolution of galaxies amounts to
3.5 $\pm$ 0.9. More recent results are those by \citet{pea03} who
finds 2.93 $\pm$ 0.85 from observations
of 30 Dor and NGC 346, and by \citet{izo06} who, from the observations
of 82 H~{\sc ii} regions, find $\Delta Y/\Delta O $ = 4.3 $\pm$ 0.7. We
have recomputed the value by Izotov et al. considering two systematic effects
not considered by them: the fraction 
of oxygen trapped in dust grains, which we estimate to be 10\% for objects of low
metallicity, and 
the increase in the $O$ abundances due to explicit taking into account the presence of temperature 
fluctuations, which for this type of H~{\sc ii} regions we estimate to be 
about 0.08 dex \citep{rel02}. From these considerations we obtain 
for the Izotov et al. sample a $\Delta Y/\Delta O  = 3.2 \pm 0.6$.

\begin{figure}[ht!]
\plotone{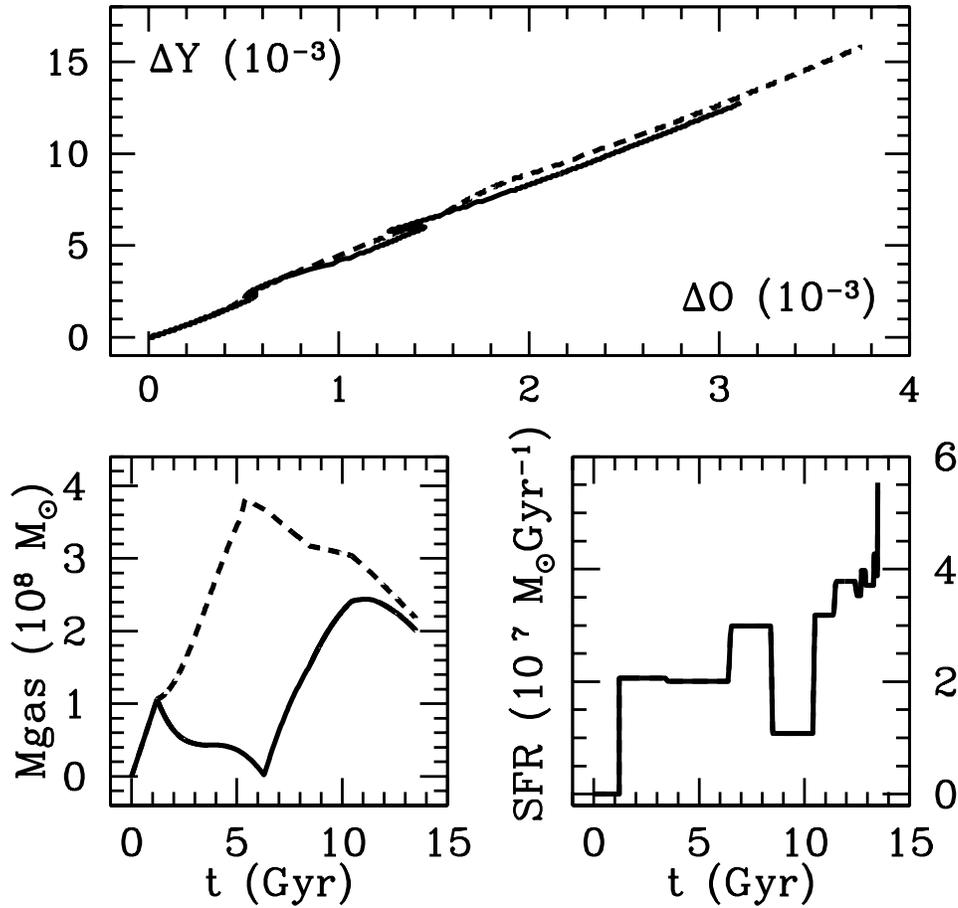}
\caption{Two chemical evolution models for NGC 6822 with different
gas infall and outflow histories, and the same star formation rate
derived from observations. The models are the 7L and 
8S (continuous and dashed lines, respectively) studied by \citet{car06}. The first panel shows the increase of
the helium and oxygen abundances relative to the primordial values,
the second panel shows the gaseous content as a function of time, which is 
widely different for the two models, and the third panel shows
the star formation history, which is the same one in both models.
}
\end{figure}

{From} chemical evolution models
of galaxies it is found that  
$\Delta Y/\Delta O$  depends on: the stellar yields,
the initial mass function, the star formation rate, 
the age, and the $O$ value of the galaxy in question.
Models with substantial outflows of O-rich material can produce large 
$\Delta Y/\Delta O $ ratios but they are ruled out by the low C/O values 
observed in irregular galaxies \citep{car95,car99,car06}.

\citet{car07} have produced chemical evolution models of the following 
types: closed box, inflow of gas, and outflow 
of gas of well-mixed material.
They find that $\Delta Y/\Delta O$ is practically constant for models with the same IMF, 
the same age, the same star formation history, and an $O$ abundance smaller 
than $\sim$ 4$\times 10^{-3}$. They
find also that $ 2.4 < \Delta Y/\Delta O < 4.1 $ for models with different star formation
histories and different values of the upper mass limit of the IMF. 
The results derived from the chemical evolution models are in very good agreement
with the observations mentioned above. 

Based on the observations and models discussed, \citet{pei07} 
adopted a value of $\Delta Y/\Delta O  = 3.3 \pm 0.7$ in the computation of the error
budget presented in Table~1 and the $Y_p$ value presented
in Table~2.

For models of galactic chemical evolution that reach values of $O > 4 \times 10^{-3}$ 
at present time, the  $\Delta Y/\Delta O $ ratio of the interstellar medium increases with
the $O$ abundance due to two effects: the helium production by low and 
intermediate mass stars and the increase of the helium yield of massive
stars due to stellar winds. 

As an example of the minor role that well mixed outflows play in the chemical
evolution of galaxies in Figure 1 we present the $\Delta Y$ versus $\Delta O$ 
behavior for two chemical evolution models of NGC 6822 \citep{car06,car07}. 
These models have the same SFH, which was derived from observations, but
are drastically different in their gas flow histories and show practically
the same $\Delta Y/\Delta O$ behavior.

To compute stellar evolution models with $O < 4 \times 10^{-3}$ we
propose to use the following relation for the initial $Y$  and $O$ abundances
\begin{equation}
Y = 0.2474 \pm 0.0028 + (3.3 \pm 0.7){ O}
\label{eq:Y}
.\end{equation}
For $O > 4 \times 10^{-3}$ \citep{car07}, $Y$ increases faster with the increase of $O$
than in the previous equation.

The ratio $\Delta Y/\Delta O$ can also be expressed in terms of $\Delta Y/\Delta Z$ if
the fraction of $O/Z$ is known. 
Based on observations of galactic and extragalactic H~{\sc ii} regions, we propose
to assume that, for models with $O < 4 \times 10^{-3}$, $O$ constitutes 55\% $\pm~5$\% of the total $Z$ value,
implying $\Delta Y/\Delta Z = 2.0 \pm 0.6$.
For $O > 4 \times 10^{-3}$ the fraction of
$Z$ due to $O$ decreases due to the increase of the C/O, N/O and Fe/O
ratios with the increase of $O$. From the chemical composition of the Orion nebula and M17
it is found that the fraction of $Z$ due to $O$ drops to about 42\% for
$0.0057 < O < 0.0082$ \citep[e.g.][and references therein]{pea03,est04,gar07}.

Previous determinations of the $\Delta Y/\Delta Z $ ratio based on models or observations
have been in the 1.0 to 6.0 range \citep[e.g.][and references therein]{pei95,fuk06}. It
is interesting to note that 25 years ago \citet{chi82} determined a value of 
$\Delta Y/\Delta Z \sim 2.0$, in excellent agreement with the recent results by 
\citet{car07}.

\section{Discussion}

In Table~2 we present some of the best $Y_p$ determinations of the last 
few years. The $Y_p$ values and their statistical errors are the ones presented
in the original papers. After the statistical error we list the systematic error
estimated by us, which depends on one or more of the following sources:
a) the change in the published emissivities of the He~{\sc i} lines \citep{por05}; 
b) the change in the published collisional excitation coefficients of the H~{\sc i} lines  \citep{and00,and02};
and 
c) the temperature structure of the H~{\sc ii} region.
Adopting the new He~{\sc i} emissivities, the $Y$ determination of individual H~{\sc ii} regions
is increased by about 0.004. The change in the H~{\sc i} collisional excitation coefficients 
goes in the same direction  \citep{pei07}, although in this case the size of this effect varies from object to object and a tailored
photoionization model for each object is needed to obtain a good estimate of it.
Both effects produce an increase in the $Y_p$ determination:
for the sample of H~{\sc ii} regions used by \citet{pei07}, the increase in $Y_p$
due to the adoption of the new He~{\sc i} emissivities amounts to about 0.0040, while the increase
due to the adoption of the new H~{\sc i} collisional excitation coefficients amounts to 0.0025.
Finally, the $Y$ value of individual H~{\sc ii} regions decreases
when the temperature structure of the H~{\sc ii} region is taken into account,
since in the self-consistent solutions for all
the observed He~{\sc i} line intensities, the lower
$T$(He~{\sc ii}) values imply higher densities, the higher densities produce a higher 
collisional contribution to the He~{\sc i} intensities, and consequently lower helium abundances.
For the objects in the sample used by \citet{pea02} and \citet{pei07}, 
this effect decreases the $Y$ determinations by amounts ranging from 0.003 to 0.009. 

The $Y_p$ determination by \citet{izo99} is affected by all three of the above sources of
systematic error; those by \citet{lal03} and \citet{izo04} are affected by sources a) and b), 
while the one by \citet{izo06} is affected by source b).
The disagreement between the $Y_p$ derived by \citet{lal03} 
with the $Y_p$ derived from WMAP under the assumption of SBBN 
implies the need for ``new physics''. The new physics needed
to reconcile the two $Y_p$ values turned out to be the ``new atomic physics'' by
\citet{and00,and02} and \citet{por05}.  With the new physics \citet{pei07}
found agreement, within the observational errors, between the observed $Y_p$
value and that derived from the WMAP results under the assumption
of SBBN.

The $Y_p$ derived by \citet{pei07} together with SBBN implies
that $\Omega_bh^2$ = $0.02054 \pm 0.00639$ \citep{ste06a,ste06b}, where $\Omega_b$ 
is the baryon closure parameter, and $h$ is the Hubble parameter. This 
value is in excellent agreement with the value derived by \citet{spe06}
from the WMAP results under the assumption of SBBN, which amounts 
to $\Omega_bh^2$ = $0.02233 \pm 0.00082$.

The comparison of the $Y_p$ derived by \citet{pei07} with the 
$Y_p$ derived by \citet{spe06} from the WMAP data together with the assumption of SBBN 
provides strong constraints for the study of non SBBN 
\citep[e.g.][and references therein]{cyb05,coc06}.

In Table~2 we also present our $Y_p$ prediction for 2010. We consider that 
in the next few years it will be possible to reduce
the statistical errors in the $Y_p$ determination to about 0.0020 by obtaining a new set of observations of
brighter and slightly O-richer H~{\sc ii} regions than the ones that have been used so far. A more extensive discussion of
the relative advantages of these H~{\sc ii} regions with respect to more metal-poor ones can be found in
\citet{pei07}.

\begin{table}[!ht]
\label{tab:pri}
\caption{Primordial helium abundance values $^a$}
{\small
\begin{center}
\begin{tabular}{lc@{ }c@{ }c}
\noalign{\smallskip}
\tableline
\tableline
\noalign{\smallskip}
{Source} & \multicolumn{3}{c}{$Y_p$} \\
\noalign{\smallskip}
\tableline 
\noalign{\smallskip}
Izotov et al. (1999), this work     & 0.2452 & $\pm$ 0.0015 & $\pm$ 0.0100  \\
Luridiana et al. (2003), this work  & 0.2391 & $\pm$ 0.0020 & $\pm$ 0.0070  \\
Izotov \& Thuan (2004), this work   & 0.2421 & $\pm$ 0.0021 & $\pm$ 0.0075  \\
Izotov et el. (2006), this work     & 0.2462 & $\pm$ 0.0025 & $\pm$ 0.0040  \\
Peimbert et al. (2007)              & 0.2474 & $\pm$ 0.0028           \\
Prediction (2010), this work        & 0.2??? & $\pm$ 0.0020           \\
Spergel et al. (2006)               & 0.2482 & $\pm$ 0.0004           \\
\noalign{\smallskip}
\tableline
\end{tabular}
\end{center}
$^a$ Direct $Y_p$ determinations based on observations of H~{\sc ii} regions, with the exception of that by Spergel et al. (2006), which is based on the baryon to photon ratio derived from WMAP and the assumption of SBBN.
}
\end{table}                                    

\section{Summary and conclusions}

In this contribution we have presented some recent determinations of $Y_p$ and discussed the reasons
underlying the differences between them. For the most recent determination, the one by \citet{pei07}, 
we have presented the error budget in terms of thirteen different sources of error. The 
$\Delta Y/\Delta O $ ratio, which enters as a crucial factor in one of the three main sources of error, 
  has been discussed in the light of recent observations and models of galactic chemical evolution.

The $Y_p$ determinations by different groups are slowly converging among them 
as systematic errors are progressively identified and corrected for. 
The need for ``new physics'' that had been suggested by recent results \citep[e.g.,][]{lal03} 
seems now to be fulfilled by new atomic physics, i.e. the atomic data
by \citet{and00,and02} and \citet{por05}. 
On the other hand, the temperature structure of
H~{\sc ii} regions is still a source of systematic error if an appropriate
scheme for temperature is not adopted.
The proper temperature to determine the helium abundance is $T$(He~{\sc ii}), derived self-consistently from the intensities of the helium lines. Adopting this temperature, the $Y_p$ value derived from H~{\sc ii} regions agrees with the $Y_p$ derived from the WMAP data under the assumption of SBBN.
On the other hand, the use of $T$(O~{\sc iii}) to determine the
$Y$ values from H~{\sc ii} regions produces  $Y_p$ values
from 0.003 to 0.006 higher than those found adopting $T$(He~{\sc ii}), 
that is $Y_p$ values more than 1$\sigma$ higher than the one predicted by the WMAP
observations combined with the assumption of SBBN.

It is important to continue the effort on the study of the physical conditions in 
H~{\sc ii} regions, this effort will permit us to lower the error on the $Y_p$ determination, 
which in turn will permit us to improve our knowledge on the possible importance of non SBBN.

\acknowledgements 
This work was partly supported by the CONACyT grant 46904 and by the Spanish 
{\it Programa Nacional de Astronom\'\i a y Astrof\'\i sica} through the project 
AYA2004-07466. VL is supported by a {\it CSIC-I3P} fellowship.


\begin{thebibliography}{}

\bibitem[Anderson et al.(2000)]{and00} 
Anderson, H., Ballance, C.~P., Badnell, N.~R., 
\& Summers, H.~P.\ 2000, 
Journal of Physics B Atomic Molecular Physics, 33, 1255 

\bibitem[Anderson et al.(2002)]{and02} 
Anderson, H., Ballance, C.~P., Badnell, N.~R., 
\& Summers, H.~P.\ 2002, 
Journal of Physics B Atomic Molecular Physics, 35, 1613 

\bibitem[\protect\citeauthoryear{Carigi, Col\'{\i}n, \& Peimbert}{Carigi
et~al.}{1999}]{car99} 
Carigi, L., Col\'{\i}n, P., \& Peimbert, M. 1999, 
\apj, 514, 787

\bibitem[Carigi et~al.(2006)]{car06}
Carigi, L., Col\'{\i}n, P., \& Peimbert, M. 2006, 
ApJ, 644, 924

\bibitem[Carigi et~al.(1995)]{car95} 
Carigi, L., Col\'{\i}n, P., Peimbert, M., \& Sarmiento, A. 1995, 
\apj, 445, 98

\bibitem[Carigi \& Peimbert(2007)]{car07} 
Carigi, L., \& Peimbert, M. 2007,
\apj, to be submitted

\bibitem[Chiosi \& Matteucci(1982)]{chi82} 
Chiosi, C., \&  Matteucci, F. 1982, 
\aap, 105, 140  

\bibitem[Coc et al.(2006)]{coc06}
Coc, A., Nunes, N. J., Olive, K. A., Usan, J.-P. \& Vangion, E. 2006,
astro-ph/0610733

\bibitem[Cyburt et al.(2005)]{cyb05}
Cyburt, R. H., Fields, B. D., Olive, K. A., \& Skillman, E. 2005,
Astroparticle Physics, 23, 313

\bibitem[Esteban et al.(2004)]{est04} 
Esteban, C., Peimbert, M., Garc\'{\i}a-Rojas, J.,  Ruiz, M. T., 
Peimbert, A., \& Rodr\'{\i}guez, M. 2004, 
\mnras, 355, 229

\bibitem[Fukugita \& Kawasaki(2006)]{fuk06} 
Fukugita, M., \&  Kawasaki, M. 2006, 
\apj, 646, 691

\bibitem[Garc\'{\i}a-Rojas et al.(2007)]{gar07}
Garc\'{\i}a-Rojas, J., Esteban, C., Peimbert, A., Rodr\'{\i}guez, 
M., Peimbert, M., \& Ruiz, M. T. 2007,
RevMexAA, in press, astro-ph/0610065
	
\bibitem[Izotov et al.(1999)]{izo99} 
Izotov, Y.~I., Chaffee, 
F.~H., Foltz, C.~B., Green, R.~F., Guseva, N.~G., \& Thuan, T.~X.\ 1999, 
\apj, 527, 757
 
\bibitem[Izotov et al.(2006)]{izo06}
Izotov, Y. I., Schaerer, D., Blecha, A., Royer, F., Guseva, N.~G.,
\& North, P. 2006,
\aap, 459, 71

\bibitem[Izotov \& Thuan(2004)]{izo04} 
Izotov, V. I., \&  Thuan, T. X. 2004, 
\apj, 602, 200

\bibitem[Luridiana(2003)]{lur03}
Luridiana, V. 2003, proceedings of the XXXVIIth Moriond
Astrophysics Meeting ``The Cosmological Model''
eds. Y. Giraud-H\'eraud, C. Magneville, J. Tran Thanh Van, The Gioi 
Publishers (Vietnam), 159  (astro-ph/0209177)

\bibitem[Luridiana et al.(2003)]{lal03} 
Luridiana, V., Peimbert, A., Peimbert, M., \& Cervi{\~n}o, M.\ 2003, 
\apj, 592, 846 

\bibitem[Olive \& Skillman(2004)]{oli04} 
Olive, K. A., \& Skillman, E. D. 2004, 
\apj, 617, 29

\bibitem[Peimbert(2003)]{pea03}
Peimbert, A. 2003, 
\apj, 584, 735

\bibitem[Peimbert et al.(2002)]{pea02}
Peimbert, A., Peimbert, M., \& Luridiana, V. 2002
\apj, 565, 668 

\bibitem[Peimbert(1995)]{pei95}
Peimbert, M. 2003, in {\it The Light Element Abundances}, ed.
P. Crane (Springer), p. 165, 1995.


\bibitem[\protect\citeauthoryear{Peimbert, Luridiana, \& Peimbert}
{Peimbert et~al.}{2007}]{pei07}
Peimbert, M., Luridiana, V., \& Peimbert, A. 2007,
\apj, submitted (astro-ph/0701580)

\bibitem[Peimbert et al.(2003)]{pei03}
Peimbert, M., Peimbert, A., Luridiana, V. \&
Ruiz, M. T. 2003, in {\it Star Formation through Time}, 
ASP Conference Series, {297}, 81

\bibitem[\protect\citeauthoryear{Peimbert, Peimbert, \& Ruiz}
{Peimbert et~al.}{2000}]{pei00}
Peimbert, M., Peimbert, A., \& Ruiz, M. T. 2000,
\apj, 541, 688

\bibitem[Porter et al.(2005)]{por05} 
Porter, R. L., Bauman, R. P., Ferland, G. J. , \& MacAdam, K. B. 2005,
\apj, 622L, 73

\bibitem[\protect\citeauthoryear{Rela{\~ n}o, Peimbert, \& Beckman}
{Rela\~no et al.}{2002}]{rel02}
Rela{\~ n}o, M., Peimbert, M., \& Beckman, J.\ 2002, 
\apj, 564, 704 


\bibitem[Spergel et al.(2006)]{spe06}
Spergel, D. N. et al. 2006, 
astro-ph/0603449

\bibitem[Steigman(2006a)]{ste06a} 
Steigman, G. 2006a,
Int. J. Mod. Phys. E, 15, 1, astro-ph/0511534

\bibitem[Steigman(2006b)]{ste06b} 
Steigman, G. 2006b,
astro-ph/0606206


\end{thebibliography}
\end{document}